\documentclass[aps,prd,twocolumn,groupedaddress,amssymb,eqsecnum,showpacs,epsfig]{revtex4}
\usepackage{graphicx}
\usepackage{bm}
\usepackage{dcolumn}
\usepackage{amsmath}
\def \lleq {\lower0.9ex\hbox{ $\buildrel < \over \sim$} ~}
\def \ggeq {\lower0.9ex\hbox{ $\buildrel > \over \sim$} ~}

\def \beq  {\begin{equation}}
\def \eeq  {\end{equation}}
\def \ber  {\begin{eqnarray}}
\def \eer  {\end{eqnarray}}

\begin{document}
\newcommand{\newc}{\newcommand}

\newc{\be}{\begin{equation}}
\newc{\ee}{\end{equation}}
\newc{\ba}{\begin{eqnarray}}
\newc{\ea}{\end{eqnarray}}
\newc{\bea}{\begin{eqnarray*}}
\newc{\eea}{\end{eqnarray*}}
\newc{\D}{\partial}
\newc{\ie}{{\it i.e.} }
\newc{\eg}{{\it e.g.} }
\newc{\etc}{{\it etc.} }
\newc{\etal}{{\it et al.}}
\newcommand{\nn}{\nonumber}
\newc{\ra}{\rightarrow}
\newc{\lra}{\leftrightarrow}
\newc{\lsim}{\buildrel{<}\over{\sim}}
\newc{\gsim}{\buildrel{>}\over{\sim}}
\title{Exploring Cosmological Expansion Parametrizations with the Gold SnIa Dataset}
\author{R. Lazkoz$^a$, S. Nesseris$^b$ and L. Perivolaropoulos$^b$}
\email{http://leandros.physics.uoi.gr} \affiliation{$^a$Fisika
Teorikoa, Zientzia eta Teknologiaren Fakultatea, Euskal Herriko
Unibertsitatea, 644 Posta Kutxatila, 48080 Bilbao, Spain
\\$^b$Department of Physics, University of Ioannina, Greece}
\date{\today}

\begin{abstract}
We use the SnIa Gold dataset to compare LCDM with 10
representative parametrizations of the recent Hubble expansion
history $H(z)$. For the comparison we use two statistical tests;
the usual $\chi_{min}^2$  and a statistic we call the p-test which
depends on both the value of $\chi_{min}^2$ and the number $n$ of
the parametrization parameters. The p-test measures the confidence
level to which the parameter values corresponding to LCDM are
excluded from the viewpoint of the parametrization tested. For
example, for a linear equation of state parametrization $w(z)=w_0
+ w_1 z$ the LCDM parameter values ($w_0=-1$, $w_1=0$) are
excluded at 75\% confidence level. We use a flat prior and
$\Omega_{0m}=0.3$. All parametrizations tested are consistent with
the Gold dataset at their best fit. According to both statistical
tests, the worst fits among the 10 parametrizations,  correspond
to the Chaplygin gas, the brane world and the Cardassian
parametrizations. The best fit is achieved by oscillating
parametrizations which can exclude the parameter values
corresponding to LCDM at 85\% confidence level. Even though this
level of significance does not provide a statistically significant
exclusion of LCDM (it is less than $2\sigma$) and does not by
itself constitute conclusive evidence for oscillations in the
cosmological expansion, when combined with similar independent
recent evidence for oscillations coming from the CMB and matter
power spectra it becomes an issue worth of further investigation.
\end{abstract}
%
%\pacs{98.80.Es,98.65.Dx,98.62.Sb}
%
\maketitle

\section{Introduction}
Converging observational evidence that appeared during the past
decade has indicated that we live in a spatially flat universe
with low matter density that is currently undergoing accelerated
cosmic expansion
\cite{Riess:2004nr,Spergel:2003cb,Readhead:2004gy,Goldstein:2002gf,Rebolo:2004vp,Tegmark:2003ud,Hawkins:2002sg}.
This accelerating expansion has been attributed to a dark
energy \cite{Sahni:2004ai} component with negative pressure which
can induce repulsive gravity and thus cause accelerated expansion.

The simplest and most obvious candidate for this dark energy is
the cosmological constant $\Lambda$ \cite{Sahni:1999gb} with
equation of state $w={p / \rho}=-1$. Such a model predicts an
expansion history of the universe which is described by an
expansion rate $H(z)$ as a function of the redshift $z$ given by
\be \label{lcdm} H^2(z;\Omega_{0m}) = \left({{{\dot a}}\over
a}\right)^2 = H_0^2 [\Omega_{0m} (1+z)^3 + (1- \Omega_{0m})] \ee
where flatness has been imposed and $\Omega_{0m}\equiv
{\rho_0}/{\rho_c}$ is the single free parameter of this simplest
data consistent parametrization (LCDM).

The most sensitive observational probe for testing this type of
parametrizations comes from distant standard candles like type Ia
supernovae (SnIa). These make it possible to start seeing the
varied effects of the universe's expansion history. The most
updated and reliable compilation of SnIa is the {\it Gold} dataset
recently relased by Riess et al. \cite{Riess:2004nr}. The authors
have compiled a catalog containing 157 SnIa with $z$ in the range
$(0.01,1.75)$ and visual absorption $A_V < 0.5$. The distance
modulus of each object has been evaluated by using a set of
calibrated methods so that the sample is homogenous in the sense
that all the SNeIa have been re-analyzed using the same technique.
Thus the resulting Hubble diagram is indeed reliable and accurate.
Even though LCDM provides the simplest parametrization consistent
with the Gold dataset it has two disadvantages which motivate the
search for other models:
\begin{itemize}
\item It requires extreme fine tuning of the value of the
cosmological constant $\Lambda$ (coincidence problem). \item It
does not provide the best possible fit to the Gold dataset.
\end{itemize}
In an effort to address these disadvantages three approaches have
been followed
\begin{itemize}
\item Assume the existence of a homogeneous time dependent scalar
field whose dynamics is determined by a specially designed
potential so that its energy comes to dominate at present and its
negative pressure plays the role of dark energy. Such fields could
either have $w>-1$ if the sign of the kinetic term is positive
(quintessence \cite{quintess}) or $w<-1$ if the sign of the
kinetic term is negative (phantom fields \cite{phantom}).
\enlargethispage{\baselineskip} \enlargethispage{\baselineskip}
\item Consider extensions of general relativity
\cite{modgrav,boisseau,Perivolaropoulos:2003we,Perivolaropoulos:2002pn,Alam:2002dv}
(motivated eg by extra dimensions) which can lead to an
accelerating expansion without any modification of the energy
momentum tensor. \item Examine arbitrary parametrizations
\cite{phant-obs2,Nesseris:2004wj} of the expansion history $H(z)$
and focus on maximizing the quality of fit to the Gold dataset.
\end{itemize}
The problem with the first two approaches which exhibit some
physical motivation is that despite their increased number of
parameters they very rarely exceed the quality of fit of LCDM.
This is not the case for the third approach. Parametrizations
which are not constrained to emerge from currently known physical
theories provide significantly better fits to the Gold dataset
than LCDM. For example it is very hard \cite{Vikman:2004dc} to
construct a physically motivated model predicting an evolution of
the dark energy equation of state $w(z)$ that crosses the line
$w=-1$ (phantom divide line).  Yet the fit to the Gold dataset
improves significantly if such crossing is allowed. The origin of
this problem could either be statistical or physical. In the
former case we are dealing with a statistical fluctuation of the
data and upcoming more detailed datasets  will become more
consistent with some of the current theories (either minimally
coupled scalar fields or modified gravity). In the later case
however, if eg the $w=-1$ crossing is verified by more detailed
data, new theories will have to be constructed. Theoretical
attempts in that direction have already started developing using
either less physically motivated approaches involving combinations
of scalar fields with positive and negative kinetic terms
\cite{Wei:2005nw,Hu:2004kh} or reconstruction of scalar tensor
theories of gravity \cite{boisseau,Perivolaropoulos:2005yv}.

In an effort to forecast and support this development it is
important to extract the maximal information from the Gold
dataset. In particular we wish to address the following  question:
`Given the number of parameters what is an arbitrary (flat and low
matter density) $H(z)$ parametrization that provides the best fit
to the Gold dataset?' Previous studies attempting to address this
type of questions \cite{Nesseris:2004wj,phant-obs2} have either
been based \cite{Nesseris:2004wj} on earlier less reliable datasets
\cite{Tonry:2003zg,Barris:2003dq} or have not been as extensive in
the search of parametrization space \cite{phant-obs2}. Here we
show that a careful and extensive search of parametrization space
can reveal interesting new features of the expansion history
hidden in the Gold dataset.

We continue and extend previous work by two of us
\cite{Nesseris:2004wj} and evaluate the quality of fit to the Gold
dataset for 10 representative parametrizations of $H(z)$. They
include physically motivated parametrizations like brane world
\cite{Alam:2002dv}, Chaplygin gas (first considered in
\cite{Kamenshchik:2001cp}, see also
\cite{Chimento:2003ta,Fabris:2001tm}) and Cardassian cosmology,
\cite{Freese:2005ff} as well as arbitrary parametrizations like a
linear fit to the equation of state ($w(z)=w_0 + w_1 z$).

The parametrizations compared have a different number of
parameters and therefore their comparison on the basis of how low
a $\chi_{min}^2$ they achieve would be biased towards
parametrizations with larger number of parameters. The reduced
$\chi_{min}^2$ ($\chi_{min}^2$ per degree of freedom) is an
improved statistic in that it has some weak dependence on the
number of parameters. We propose however an alternative statistic,
we call `p-test' which avoids the above bias and is more
appropriate for the comparison of parametrizations with different
number of parameters. This statistic along with the general method
used for evaluating the quality of fit to the Gold dataset of the
parametrizations considered is described in the next section. In
section III we present our results for the best fits and discuss
their implications and common features.

\section{Fitting Parametrizations to the Gold dataset}
Given a parametrization $H(z;a_1,...,a_n)$ depending on $n$
parameters we can obtain the corresponding Hubble free luminosity
distance \be D_L^{th} (z;a_1,...,a_n)= (1+z) \int_0^z
dz'\frac{H_0}{H(z';a_1,...,a_n)} \label{dlth1} \ee Using the
maximum likelihood technique \cite{press92} we can find the
goodness of fit to the corresponding observed $D_L^{obs} (z_i)$
$(i=1,...,157)$ coming from the SnIa of the Gold dataset
\cite{Riess:2004nr}. The observational data of the gold dataset
are presented as the apparent magnitudes $m(z)$ of the SnIa with
the corresponding redshifts $z$ and $1\sigma$ errors
$\sigma_{m(z)}$. The apparent magnitude is connected to $D_L (z)$
as \be m(z;a_1,...,a_n)={\bar M} (M,H_0) + 5 log_{10} (D_L
(z;a_1,...,a_n)) \label{mdl} \ee where ${\bar M}$ is the magnitude
zero point offset and depends on the absolute magnitude $M$ and on
the present Hubble parameter $H_0$ as \be {\bar M} = M + 5
log_{10}(\frac{c\; H_0^{-1}}{Mpc}) + 25 \label{barm} \ee The
goodness of fit corresponding to any set of parameters
$a_1,...,a_n$ is determined by the probability distribution of
$a_1,...,a_n$ \ie \be P({\bar M}, a_1,...,a_n)= {\cal N} e^{-
\chi^2({\bar M},a_1,...,a_n)/2} \label{prob1} \ee where  \be
\chi^2 ({\bar M},a_1,...,a_n)= \sum_{i=1}^{157}
\frac{(m^{obs}(z_i) - m^{th}(z_i;{\bar
M},a_1,...,a_n))^2}{\sigma_{m^{obs}(z_i)}^2} \label{chi2} \ee and
${\cal N}$ is a normalization factor. If prior information is
known on some of the parameters $a_1,...,a_n$ then we can either
fix the known parameters using the prior information or
`marginalize', i.e. average the probability distribution
(\ref{prob1}) around the known value of the parameters with an
appropriate `prior' probability distribution.
\enlargethispage{\baselineskip} \enlargethispage{\baselineskip}
\hspace{-0.18cm}The parameters ${\bar a}_1,...,{\bar a}_n$ that
minimize the $\chi^2$ expression (\ref{chi2}) are the most
probable parameter values (the 'best fit') and the corresponding
$\chi^2({\bar a}_1,...,{\bar a}_n)\equiv \chi_{min}^2$ gives an
indication of the quality of fit for the given parametrization:
the smaller $\chi_{min}^2$ is, the better the parametrization.

The minimization with respect to the parameter $\bar{M}$ can be
made trivially by expanding \cite{sfay} the $\chi^2$ of equation
(\ref{chi2}) with respect to $\bar{M}$ as \be  \chi^2 (a_1,..,a_n)
= A - 2 {\bar M} B  + {\bar M}^2 C  \label{chi2bm} \ee where \ba
A(a_1,..,a_n)&=&\sum_{i=1}^{157} \frac{(m^{obs}(z_i) - m^{th}(z_i
;{\bar
M}=0,a_1,..,a_n))^2}{\sigma_{m^{obs}(z_i)}^2} \label{bb} \nn \\
B(a_1,..,a_n)&=&\sum_{i=1}^{157} \frac{(m^{obs}(z_i) - m^{th}(z_i
;{\bar
M}=0,a_1,..,a_n))}{\sigma_{m^{obs}(z_i)}^2} \label{bb} \nn \\
C&=&\sum_{i=1}^{157}\frac{1}{\sigma_{m^{obs}(z_i)}^2 } \label{cc}
\ea Equation (\ref{chi2bm}) has a minimum for ${\bar M}={B}/{C}$
at \be {\tilde\chi}^2(a_1,...,a_n)=A(a_1,...,a_n)-
\frac{B(a_1,...,a_n)^2}{C} \label{chi2min1}\ee Thus instead of
minimizing $\chi^2({\bar M},a_1,...,a_n)$ we can minimize
${\tilde\chi}^2(a_1,...,a_n)$ which is independent of ${\bar M}$.
Obviously $\chi_{min}^2={\tilde\chi}_{min}^2$. Alternatively we
could have marginalized over the nuisance parameter ${\bar M}$
thus obtaining
\cite{Nesseris:2004wj,Perivolaropoulos:2004yr,DiPietro:2002cz} \be
{\tilde\chi}^2(a_1,...,a_n)=A(a_1,...,a_n)-
\frac{B(a_1,...,a_n)^2}{C} + \ln (C/2\pi) \label{chi2min2}\ee to
be minimized with respect to $a_1,...,a_n$. In our analysis we
consider the ${\tilde\chi}^2(a_1,...,a_n)$ of equation
(\ref{chi2min1}) which is already minimized with respect to ${\bar
M}$. In what follows we omit the symbol tilde (${\tilde .}$) for
simplicity.

Clearly the value of $\chi_{min}^2$ depends on the parametrization
used, and it can decrease arbitrarily close to 0 for large enough
number of parameters $n$. Therefore, the value of $\chi_{min}^2$
by itself is not a particularly useful measure of the quality of a
given parametrization. It is biased toward parametrizations with
larger number of parameters and is therefore useful only in
comparing parametrizations with the same number of parameters $n$.
Notice however that the reduced $\chi_{min}^2$ ($\chi_{min}^2$ per
degree of freedom) has some weak dependence on the number of
parameters and can be an alternative when comparing the quality of
fit of parametrizations with different number of parameters.

How can we compare parametrizations with different number of
parameters in a more efficient way? This can be achieved by the
following procedure: Consider two parametrizations of the
expansion history with different number of parameters $H_1
(z;a_1,...,a_n)$ and $H_2 (z;b_1,...,b_m)$ and assume that for the
parameter values $a_1^L,...,a_n^L$ and $b_1^L,...,b_m^L$ both
parametrizations reduce to a common form eg LCDM \ie \ba H_1^2
(z;a_1^L,...,a_n^L)&=&H_2^2 (z;b_1^L,...,b_m^L)=\\
\nn =H_L^2(z;\Omega_{0m})&\equiv & H_0^2 [\Omega_{0m} (1+z)^3 +
(1- \Omega_{0m})] \label{comf} \ea \enlargethispage{\baselineskip}
\enlargethispage{\baselineskip} \enlargethispage{\baselineskip}
\enlargethispage{\baselineskip}
 Let also $\chi_{1min}^2$ and
$\chi_{2min}^2$ be the minimum values of $\chi^2$ for each
parametrization and $\chi_L^2$ the value of $\chi^2$ corresponding
to LCDM. Since $H_L$ has a single parameter ($\Omega_{0m}$) which
is constrained by prior information (other observations) it is
natural to expect that $\chi_{1min}^2$ and $\chi_{2min}^2$ will be
smaller than $\chi_L^2$. Define also $\Delta \chi_1^2\equiv
\chi_1^2 - \chi_{1min}^2$, $\Delta \chi_2^2\equiv \chi_2^2 -
\chi_{2min}^2$, $\Delta \chi_{1L}^2\equiv \chi_L^2 -
\chi_{1min}^2$ and $\Delta \chi_{2L}^2\equiv \chi_L^2 -
\chi_{2min}^2$. Now  $\Delta \chi_1^2$ and $\Delta \chi_2^2$  are
random variables that obey a $\chi^2$ probability distribution
with $n$ and $m$ degrees of freedom respectively. Therefore the
probability that $\Delta \chi_1^2$ does not exceed the value
$\Delta \chi_{1L}^2$ is \ba P_{1<} (\Delta \chi_{1L}^2,n)&=&
\frac{1}{2\Gamma(n/2)}
\int_0^{\Delta\chi_{1L}^2/2}\left(\frac{x}{2}\right)^{\frac{n}{2}-1}e^{-\frac{x}{2}}dx= \nn \\
 &=& 1-\frac{\Gamma({n}/{2},{\Delta
\chi_{1L}^2}/{2})}{\Gamma({n}/{2})} \label{prex} \ea and similarly
for $\Delta \chi_2^2$. In eq. (\ref{prex}) $\Gamma({n}/{2},{\Delta
\chi_{1L}^2}/{2})$ is the incomplete $\Gamma$ function. For $n=1$
and $\Delta \chi_{1L}^2=1$ we find $P_{1<}(1,1)=0.683$
corresponding to $1\sigma$ as expected from well known tables
\cite{press92}.

 Therefore $P_{1<}
(\Delta \chi_{1L}^2,n)$ expresses the probability that given the
$n$ parameter parametrization $H_1 (z;a_1,...,a_n)$, the true
values of parameters $a_1,...,a_n$ produce a $\chi_1^2$ that is
less than the value of $\chi_{1L}^2$ corresponding to LCDM and
therefore LCDM is not realized in nature. In other words it is the
confidence level to which LCDM is excluded, from the viewpoint of
the parametrization $H_1 (z;a_1,...,a_n)$. It is easy to see that
$P_{1<} (\Delta \chi_{1L}^2,n)$ increases with $\Delta
\chi_{1L}^2$ but decreases with $n$ \ie a parametrization with
many parameters has more difficulty to exclude LCDM unless it can
provide a very low $\chi_{1min}^2$ (large $\Delta \chi_{1L}^2$).
Thus $P_{1<} (\Delta \chi_{1L}^2,n)$ provides a quantitative
measure of the quality of a given parametrization $H_1
(z;a_1,...,a_n)$ which allows comparison with a different
parametrization $H_2 (z;b_1,...,b_m)$. We call this measure
`p-test' and it will be used in what follows to compare the
quality of the parametrizations considered.

Another way of comparison of the cosmological models is by using
Bayessian theory. In simple terms this is done by forming the
so-called Bayes factor\cite{John:2002gg} $B_{ij}$, where \be
B_{ij} \equiv \frac{L(M_i)}{L(M_j)} \ee and $L(M_i)$ denotes the
probability $p(D|M_i)$ (called likelihood for the model $M_i$) to
obtain the data D if the model $M_i$ is the true one. Generally,
$L(M_i)$ is defined as: \be L(M_i)\equiv p(D|M_i) =\int da \cdot
p(a| M_i) {\cal L}_i(a)\ee for models with one free parameter and
where $p(a| M_i)$ is the prior probability for the parameter $a$.
Also, ${\cal L}_i(a)$ is the likelihood for the parameter $a$ in
the model and \be {\cal L}_i(a)\equiv e^{-\chi^2(a)/2} \ee

In the case that $a$ has flat prior probabilities, that is we have
no prior information on $a$ besides that it lies in some range
$[a,a+\Delta a]$ then $p(a|M_i)=\frac{1}{\Delta a}$ and \be
L(M_i)=\frac{1}{\Delta a}\int_a^{a+\Delta a} da e^{-\chi^2(a)/2}
\ee Of course, all this can be generalized for models having more
than one parameters.

The interpretation of the Bayes factor $B_{ij}$ is
that\cite{John:2002gg} when $1<B_{ij}<3$ there is evidence against
$M_j$ when compared with $M_i$, but it is only worth a bare
mention. When $3<B_{ij}<20$ the evidence against $M_j$ is definite
but not strong. For $20<B_{ij}<150$ the evidence is strong and for
$B_{ij}>150$ it is very strong.

In the next section we will use the Bayes factor $B_{ij}$ to
compare various parametrizations to LCDM.

\section{Results-Discussion}
In this section we utilize the p-test to rank 10 representative
parametrizations of $H(z)$. The comparison of the parametrizations
considered is shown in Table I. This is a representative sample
among the many more parametrizations we considered but we do not
present here for clarity reasons. The rank in Table I is according
to the p-test statistic indicated in the third column. A similar
rank is obtained using the $\chi_{min}^2/dof$ which is also
sensitive to the number of parameters and shown in the last column
of Table I. Notice however that the rank according to
$\chi_{min}^2$ (shown on the fourth column of Table I) which is
insensitive to the number of parameters is somewhat different.

The model name initials of the first column correspond to the
following: `OA Var (1)' is an oscillating ansatz $H(z)$ with
amplitude decreasing with time like $(1+z)^3$. `OA (2)' is a
similar ansatz but with constant oscillation amplitude. `LA (3)'
is an ansatz used by Linder \cite{Linder:2002dt} (but proposed
earlier in \cite{Chevallier:2000qy}) and is derived from an
equation of state \enlargethispage{\baselineskip}
\enlargethispage{\baselineskip} \enlargethispage{\baselineskip}
\enlargethispage{\baselineskip} \hspace{-0.28cm} parameter of the
form \be w(z)=w_0+w_1 \frac{z}{1+z} \label{lda} \ee interpolating
between two constant values of $w(z)$. `P2' is a quadratic
polynomial of $H(z)$ also discussed in Ref. \cite{Alam:2004jy}.
`Linear' is an ansatz derived by demanding a linear form of the
equation of state \be w(z)=w_0+w_1 z \label{lda} \ee `P3' is a
cubic polynomial of $H(z)$. Here we present the deeper minimum of
$\chi^2$ for P3 which shows oscillating behavior. We have found
another minimum less deep which is very similar to the minimum of
P2. `CA' is an ansatz based on the generalized cardassian
cosmology \cite{Freese:2005ff}. `Quiess' corresponds to dark
energy with constant equation of state $w$. `MCG' is a modified
form \cite{Chimento:2003ta} of the Chaplygin gas
\cite{Fabris:2001tm}. We have tested several variants of this
ansatz but they all have a fit that is marginally distinguishable
from LCDM. `Brane2' is an ansatz motivated from brane world
cosmology \cite{Alam:2002dv}. It appears in two variants which
involve a change of sign in the square root appearing in $H^2(z)$
but we find no significant difference in the quality of fit
between the two variants (`Brane1' and `Brane2'); they are both
practically indistinguishable from LCDM.

\newpage
%\begin{widetext}
\vspace{-2pt}
\begin{table}[p!]  %%%%%%%%% Table of EOS parametrizations %%%%%%%%%%%
\begin{center}
\begin{minipage}{0.88\textwidth}
\caption{A comparison of the parametrizations used in the
literature. In all cases we have assumed priors corresponding to
flatness and $\Omega_{0m}=0.3$. \label{table1}}
\end{minipage}\\
\begin{tabular}{cccccc}
\hline
\hline\\
\bf{Model} & $\bf{ H^2(z)}$, ($\bf{\Omega_{0m}=0.3}$)&\bf{p-test}
& $\bf{\chi_{min}^2}$  & {\bf Best fit parameters}&$\bf{\chi_{min}^2}/dof $\\
\\\hline \vspace{-5pt}\\
& $H^2 (z)=H_0^2 \ [\Omega_{0m} (1+z)^3 +1-\Omega_{0m}+ $  &    &  &$a_1=0.13\pm 0.07 $\\
OA Var. (1)  & $a_1 (1+z)^3[\cos(a_2 z+a_3\pi)-\cos(a_3\pi)]]$ & 0.85 & 171.733 & $a_2=6.83\pm 1.61 $& 1.115 \\
&&&&$a_3=4.57\pm 0.07 $\\
\\
& $H^2 (z)=H_0^2 \ [\Omega_{0m} (1+z)^3 + a_1\cos(a_2z+a_3\pi)+$  &    &  &$\;\;a_1=-0.30\pm 0.17 $,\;\;\\
OA (2)  & $(1-a_1\cos(a_3\pi)-\Omega_{0m})]$ & 0.81 & 172.368 & $a_2=\;\;\;\;6.34\pm 3.19 $ & 1.119 \\
&&&&$a_3=\hspace{1pt}-0.37\pm 0.14 $\\\\
& $H^2 (z)=H_0^2 [ \Omega_{0m} (1+z)^3+$ & &  & $w_0=-1.58\pm 0.33
$
\\
LA (3)&$+(1-\Omega_{0m})(1+z)^{3(1+w_0+w_1)}e^{3w_1[1/(1+z)-1]}]$& 0.79 & 173.928 & $w_1=\;\;\;3.29\pm 1.76 $&1.122 \\\\
\\& $H^2 (z)=H_0^2 \{\Omega_{0m} (1+z)^3 + a_1(1+z)+ $  &  &  &$a_1=-4.16\pm 2.53$
\\
P2 (4)  & $a_2(1+z)^2+(1-\Omega_{0m}-a_1-a_2)\}$& 0.78 & 174.207 & $a_2=\;\;\;1.67\pm 1.03 $&1.124\\\\
\\
& $H^2 (z)=H_0^2 [\Omega_{0m} (1+z)^3 + $  &  &  &$w_0=-1.40\pm
0.25$ \\
Linear (5)  & $(1-\Omega_{0m})(1+z)^{3(1+w_0-w_1)}e^{3w_1 z}] $& 0.75 & 174.365 & $w_1=\;\;\;1.66\pm 0.92 $&1.125\\\\
\\
& $H^2 (z)=H_0^2 [\Omega_{0m}(1+z)^3 + a_1(1+z)+a_2(1+z)^2+$  &  &
&$a_1=-21.79\pm 17.92$\\
P3 (6)  & $a_3 (1+z)^3+(1-\Omega_{0m}-a_1-a_2-a_3)]$& 0.74 & 173.155 & $a_2=\hspace{7pt}14.75\pm 13.07$&1.124\\
&&&&$a_3=\hspace{7pt}-3.13\pm 3.07$\\\\
& $H^2 (z)=H_0^2[\Omega_{0m} (1+z)^3 + $  &  &
&$\hspace{3pt}q=0.0058\pm 0.0014$\\
CA (7)  & $ \Omega_{0m}(1+z)^3((1+(\Omega_{0m}^{-q}-1)(1+z)^{3(n-q)})^{\frac{1}{q}}-1)]$& 0.50 & 175.758 & $\hspace{-16pt}n=-119\pm 40$&1.134\\
\\\\
& $H^2 (z)=H_0^2 [\Omega_{0m} (1+z)^3 + $  &   &   &\\
Quiess (8)& $+ (1-\Omega_{0m})(1+z)^{3(1+w)}]$&0.15&177.091& $w=-1.02\pm 0.11$&1.135\\
\\\\
& $H^2 (z)=H_0^2 [\Omega_{0m} (1+z)^3 + $  &  &
&$\hspace{9pt}a_1=0.9992 _{-0.0060}^{+0.0008}$\\
GCG (9)  & $(1-\Omega_{0m})(a_1+(1-a_1)(1+z)^3)^{w}]$& 0.031 &
177.064 & $\hspace{-6pt}w=18.13\pm 4.95$ &1.142\\
\\\\
& $H^2 (z)=H_0^2 [\Omega_{0m} (1+z)^3 - \sqrt{a_1 + a_2 (1 + z)^3}
+ $  &  &  &$a_1=29.08\pm 7.30$\\
Brane 2 (10)  & $(1 - \Omega_{0m} + \sqrt{a_1 + a_2})]$& 0.027 & 177.071 & $\hspace{12pt}a_2=-0.097\pm 0.459$&1.142\\
\\\\\\
LCDM & $H^2 (z)=H_0^2 [\Omega_{0m} (1+z)^3 + 1- \Omega_{0m}]$ & -
& 177.072  &$\Omega_{0m}=0.31\pm 0.04$&1.135\\\\
 \hline \hline
\end{tabular}
\end{center}
\end{table}
%\end{widetext}
\newpage

 In all cases we have assumed priors corresponding to flatness and $\Omega_{0m}=0.3$. For a
prior $\Omega_{0m}=0.25$ we have found no significant changes in
the results of Table I. The following are noteworthy features of
Table I:
\begin{itemize}
\item The ansatz giving the best fit to the Gold dataset is an
oscillating parametrization where the amplitude of the
oscillations decreases like the third power of the scale factor.
This type of decreasing amplitude oscillations can be induced by
an oscillating scalar field minimally or non-minimally coupled
(\eg the radion
\cite{Perivolaropoulos:2003we,Perivolaropoulos:2002pn}). With
respect to this parametrization the simplest data consistent
parametrization (LCDM) is excluded at the $85\%$ confidence level.
This level of significance is clearly not enough to disfavor LCDM
but it may provide a hint for interesting better fits to the data.
\item The cubic polynomial parametrization P3 which provides a
fairly good fit to the Gold dataset (the third best
$\chi_{min}^2$) also shows indications of $H(z)$ oscillations at
its best fit despite the fact that no oscillating functions are
built in the parametrization. We have seen a similar behavior in
other power law parametrizations even though we have not included
those in Table I to avoid confusion. \item Physically motivated
parametrizations like the Chaplygin gas, brane-world models and
Cardassian cosmology provide marginally better fits than LCDM and
are disfavored by our test due to their larger number of
parameters compared to LCDM.  \item Parametrizations that allow
$w(z=0)<-1$ and crossing of the phantom divide line $w=-1$ produce
significantly better fits to the Gold dataset.
\end{itemize}

In Table II, we consider the same parametrizations, as in Table I,
but now we have marginalized $\chi^2 $ over $\Omega_m $. We have
used uniform marginalization in the range $\Omega_m \in
[0.20,0.34]$. We also show the Bayes factor for each model against
LCDM, calculated with the techniques described in the previous
section and assuming uniform probability distribution of the
parameters within their $1\sigma$ range.

Also, as expected, after marginalizing $\chi^2 $ over $\Omega_m $,
the error bars of the best-fit parameters have increased. For
example, for the P2 model, in the case of the $\Omega_m =0.3$
prior, we found $a_1=-4.16 \pm 2.53$ and $a_2=1.67 \pm 1.03$ , but
after marginalizing $a_1=-4.45 \pm 2.81$ and $a_2=1.87 \pm 1.28$.
In the same fashion, for the LA model we found in the first case
$w_0=-1.58\pm 0.33$ and $w_1=3.29 \pm 1.76$ and in the latter
$w_0=-1.44\pm 0.44$ and $w_1=3.09 \pm 1.87$.

In Figs 1 and 2 we show the best fit $w(z)$ and $H(z)$ for 6
representative parametrizations of Table I.

Our result for the improved fit of oscillating parametrizations is
consistent with a previous finding by two of us
\cite{Nesseris:2004wj} using an earlier SnIa dataset
\cite{Tonry:2003zg,Barris:2003dq}. Clearly, even though LCDM is
disfavored at the 85\% confidence level with respect to the
oscillating parametrization this is not enough to offer conclusive
evidence for the existence of oscillations in the cosmological
expansion.

\clearpage
%\begin{widetext}
\vspace{-2pt}
\begin{table}[h!]  %%%%%%%%% Table of EOS parametrizations %%%%%%%%%%%
\begin{center}
\begin{minipage}{0.88\textwidth}
\caption{Like the previous table but now with $\chi^2$
marginalized over $\Omega_{m}$ in the range $\Omega_m \in
[0.20,0.34]$. Also, it contains the Bayes Factor $B_{ij}$, of each
model against LCDM. The Bayes factor is not shown for the Brane 2
and the P2 parametrizations because the marginalization integrals
were ill defined in part of the parameter space ($H(z)$ becomes
complex for some parameter values).\label{table2}}
\end{minipage}\\
\begin{tabular}{ccccc}
\hline
\hline\\
\bf{Model} & $\bf{ H^2(z)}$ &\bf{Bayes Factor $B_{ij}$}
& $\bf{\chi_{min}^2}$  & $\bf{\chi_{min}^2}/dof $\\
\\\hline \vspace{-5pt}\\
& $H^2 (z)=H_0^2 \ [\Omega_{0m} (1+z)^3 +1-\Omega_{0m}+ $  &    &  \\
OA Var. (1)  & $a_1 (1+z)^3[\cos(a_2 z+a_3\pi)-\cos(a_3\pi)]]$ & 9.03 & 176.442 & 1.146 \\
\\
& $H^2 (z)=H_0^2 \ [\Omega_{0m} (1+z)^3 + a_1\cos(a_2z+a_3\pi)+$  &    &  \\
OA (2)  & $(1-a_1\cos(a_3\pi)-\Omega_{0m})]$ & 2.56 & 177.029  & 1.150 \\
&&&&\\\\
& $H^2 (z)=H_0^2 [ \Omega_{0m} (1+z)^3+$ & &  &
\\
LA (3)&$+(1-\Omega_{0m})(1+z)^{3(1+w_0+w_1)}e^{3w_1[1/(1+z)-1]}]$& 2.63 & 178.676 &1.153 \\\\
\\& $H^2 (z)=H_0^2 \{\Omega_{0m} (1+z)^3 + a_1(1+z)+ $  &  &
\\
P2 (4)  & $a_2(1+z)^2+(1-\Omega_{0m}-a_1-a_2)\}$& - & 178.874 &1.154\\\\
\\
& $H^2 (z)=H_0^2 [\Omega_{0m} (1+z)^3 + $  &  &  \\
Linear (5)  & $(1-\Omega_{0m})(1+z)^{3(1+w_0-w_1)}e^{3w_1 z}] $& 2.50 & 179.173 &1.156\\\\
\\
& $H^2 (z)=H_0^2 [\Omega_{0m}(1+z)^3 + a_1(1+z)+a_2(1+z)^2+$  &  &\\
P3 (6)  & $a_3 (1+z)^3+(1-\Omega_{0m}-a_1-a_2-a_3)]$& 3.68 & 177.859 &1.155\\&&&\\\\
& $H^2 (z)=H_0^2[\Omega_{0m} (1+z)^3 + $  &  &\\
CA (7)  & $ \Omega_{0m}(1+z)^3((1+(\Omega_{0m}^{-q}-1)(1+z)^{3(n-q)})^{\frac{1}{q}}-1)]$& 0.78 & 182.308 &1.176\\
\\\\
& $H^2 (z)=H_0^2 [\Omega_{0m} (1+z)^3 + $  &   &   &\\
Quiess (8)& $+ (1-\Omega_{0m})(1+z)^{3(1+w)}]$&0.98&181.614&1.164\\
\\\\
& $H^2 (z)=H_0^2 [\Omega_{0m} (1+z)^3 + $  &  &\\
GCG (9)  & $(1-\Omega_{0m})(a_1+(1-a_1)(1+z)^3)^{w}]$& 1.26 &181.394 &1.177\\\\\\
& $H^2 (z)=H_0^2 [\Omega_{0m} (1+z)^3 - \sqrt{a_1 + a_2 (1 + z)^3}
+ $  &  &  \\
Brane 2 (10)  & $(1 - \Omega_{0m} + \sqrt{a_1 + a_2})]$& - & 182.326 & 1.176\\\\\\\\
LCDM & $H^2 (z)=H_0^2 [\Omega_{0m} (1+z)^3 + 1- \Omega_{0m}]$ & 1 & 182.326  &1.161\\\\
 \hline \hline
\end{tabular}
\end{center}
\end{table}
%\end{widetext}
\newpage
\clearpage

However, it is recently becoming clear that the possibility of
cosmological expansion oscillations is favored by a number of
independent additional factors, observational and theoretical. The
theoretical factors include the following:
\begin{enumerate}
\item The eternal acceleration predicted by LCDM and by most other
parametrizations creates problems in string theory where
asymptotic flatness of space-time is required \cite{1}.
Theoretical models (using eg special quintessence potentials or
generalizations of the Einstein action and extra dimensions
\cite{Collins:2001ic}) have been proposed \cite{Rubano:2003er} to
cure this problem by cosmological expansion oscillations.
Unfortunately, these models\cite{Rubano:2003er} do not predict
crossing of the $w=-1$ as indicated by the data because they are
based on minimally coupled scalar fields. \item A periodic
expansion rate of the universe could resolve the cosmic
coincidence problem. In an oscillating expansion setup, the
present period of acceleration would be nothing surprising; it
would merely be one of the many other accelerating periods that
also occurred in the past \cite{Dodelson:2001fq,2}. \item Models
with extra dimensions require a mechanism to stabilize the size of
extra dimensions (the radion field). However, since the radion
field also couples to the redshifting matter density (it behaves
like a Brans-Dicke scalar) its equilibrium point is slowly
shifting with time and radion oscillations are generically
excited. The energy density of these oscillations is dropping
approximately like the third power of the scale factor
\cite{Perivolaropoulos:2002pn}. It is exactly this power that
gives the best fit of the SnIa data! The constraints imposed on
such low frequency oscillations by local gravity tests can be
relaxed by taking \enlargethispage{\baselineskip}
\enlargethispage{\baselineskip} \hspace{-0.25cm}into account the
increased average matter density in the neighborhood of the solar
system which can significantly increase the local effective mass
of the radion \cite{Brax:2004ym}.
\end{enumerate}
The observational factors favoring oscillating expansion include
the following:
\begin{enumerate}
\item Superimposed oscillations \cite{Martin:2004yi} or glitches
\cite{Hunt:2004vt} in the CMB multipole moments $C_l$ can improve
the fit to the first year Wilkinson Microwave Anisotropies Probe
(WMAP) data. The corresponding drop in the $\chi^2$ for superposed
oscillations was found to be $\Delta \chi^2\simeq 10$ which is
statistically significant \cite{Martin:2004yi}. These oscillations
were attributed to new physics taking place at the inflationary
phase reflecting on
\enlargethispage{\baselineskip}
\hspace{-0.2cm}the produced fluctuations power spectrum.
However, small amplitude oscillations of the cosmological
expansion could also induce similar effects with remnants in the
present day expansion rate.

\item The 128 Mpc periodicity in the galaxy redshift distribution
was first observed in pencil redshift surveys and confirmed more
recently by large scale redshift surveys
\cite{Bajan:2003yv,Broadhurst:1990be,Einasto:1997md,a8}. This
periodic distribution could either be attributed to specific
features of the matter spectrum (coming eg from the acoustic
oscillations \cite{Eisenstein:2005su} at recombination or from
exotic physics), or to oscillating cosmological expansion
\cite{a11,a13a}. Even though the origin of this periodicity
remains an unresolved issue it should be pointed out that a recent
study has indicated that the probability that such periodicity
emerges statistically in a LCDM cosmology is less than 0.1\%
\cite{Gonzalez:2000qy,Yoshida:2000zq}.
\end{enumerate}
In conclusion, we have performed an extensive comparative study in
parametrization space and have evaluated the quality of fit of
several $H(z)$ parametrizations to the Gold dataset. In comparing
the quality of fit we have used two statistics: the value of
$\chi^2$ at its minimum ($\chi_{min}^2$) which is independent of
the parametrization number of parameters and the p-test
($P_<(\chi_{min}^2,n)$) which evaluates the quality of fit with
respect to LCDM and depends on both $\chi_{min}^2$ and the number
of parametrization parameters $n$. According to both statistics
the best fit to the Gold dataset is achieved by an oscillating
parametrization with oscillating amplitude decreasing like the
inverse cubic power of the scale factor. Even a cubic polynomial
parametrization is showing oscillating behavior at best fit. We
stress however that the statistical significance of the improved
fit of oscillating and other parametrizations we found is at less
than $2\sigma$ level and is therefore not enough to exclude the
LCDM parametrization. It does provide however a hint towards
parametrizations that are clearly more probable than LCDM to be
realized in nature on the basis of the Gold dataset.

We have also briefly reviewed the theoretical and observational
evidence that has been accumulating during the recent years in
support of such oscillations. The particular type of oscillations
that are providing the best fit to the equation of state $w(z)$
exhibit crossings of the $w=-1$ divide line. These crossings can
not be reproduced in any single field quintessence or phantom
model and probably require non-minimally coupled theories for
their physical interpretation.

The mathematica file with the numerical analysis of the paper can
be found at http://leandros.physics.uoi.gr/goldfit.htm

\section*{Acknowledgements}
 The work of S.N. and L.P.  was supported by the program
PYTHAGORAS-1 of the Operational Program for Education and Initial
Vocational Training of the Hellenic Ministry of Education under
the Community Support Framework and the European Social Fund. The
work of R.L. was supported by the Spanish Ministry of Education
and Culture  through research grant  FIS2004-01626. We would also
like to thank S.~P.~Lee for pointing out some typographic errors
on the manuscript.

\clearpage
%\begin{minipage}{\textwidth}
$\,$
\begin{figure}[t!]
\includegraphics[width=10cm,height=16cm,angle=-90]{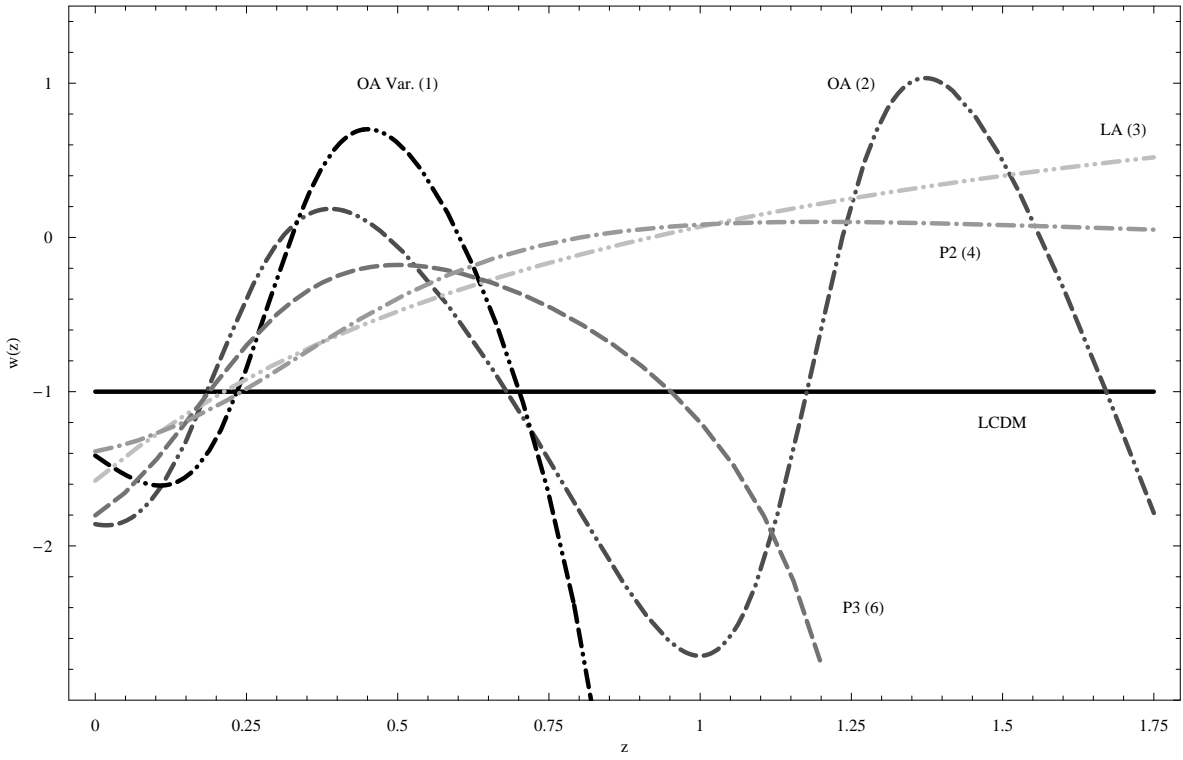}
\begin{minipage}{\textwidth}
\caption{The plot of w(z) for some representative parametrizations.}
\end{minipage}
\end{figure}

\begin{figure}[b!]
\includegraphics[width=10cm,height=16cm,angle=-90]{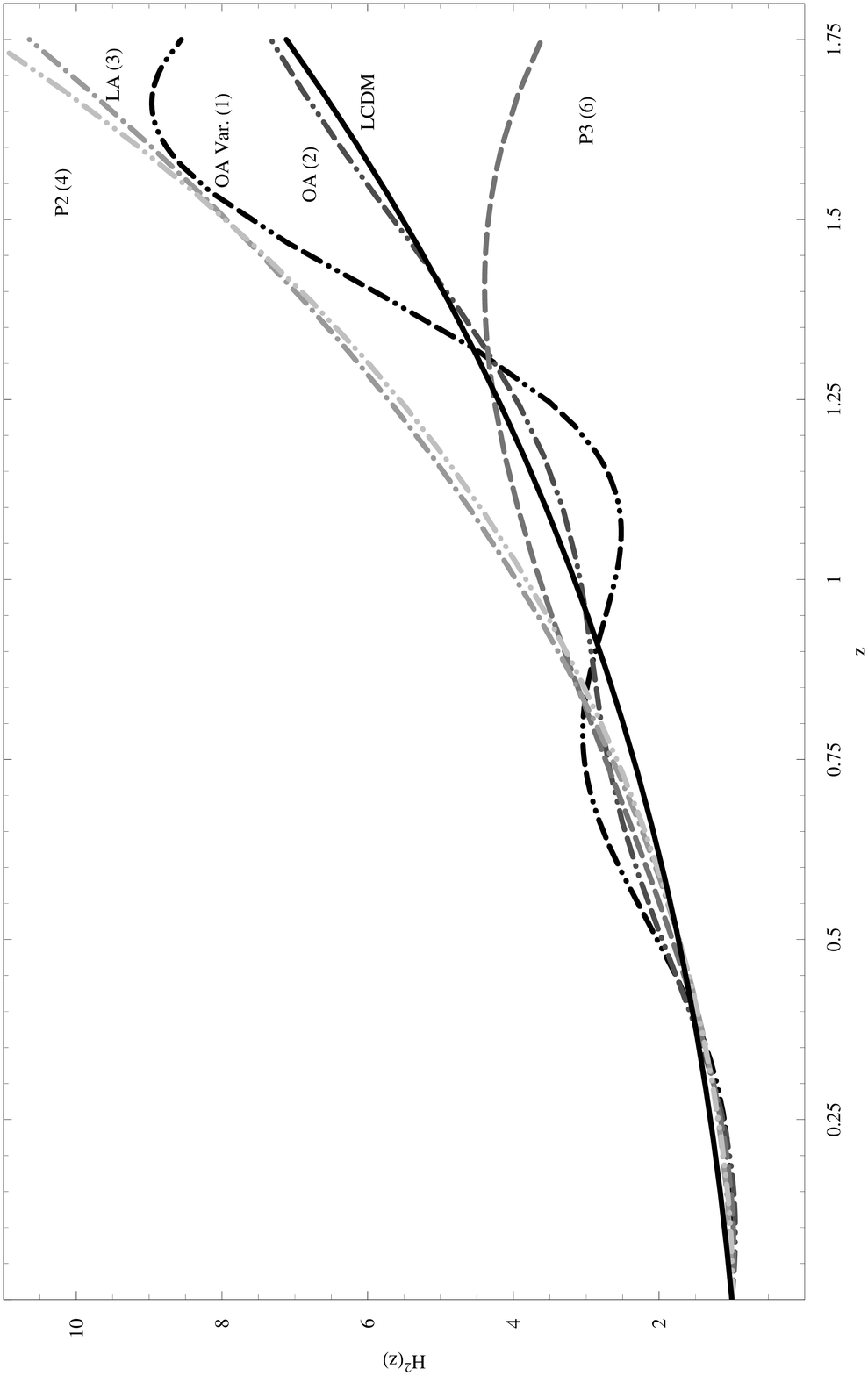}
\begin{minipage}{\textwidth}\caption{The plot of $H(z)$ for some representative parametrizations.}
\end{minipage}
\end{figure}
%\end{minipage}

\end{document}